\documentclass[%
 reprint, 
superscriptaddress,
amsmath,amssymb,
 aps,prl
]{revtex4-1}

\usepackage{graphicx}
\usepackage{bm}
\usepackage{hyperref}

\newcommand{\DM}{Dzyaloshinskii-Moriya}

\newcommand{\Neel}{N\'{e}el }
\newcommand{\rsk}{r_\mathrm{Sk}}
\newcommand{\Keff}{K_\mathrm{eff}}

\usepackage{xcolor}

\begin{document}

\title{Stray field signatures of \Neel textured skyrmions in Ir/Fe/Co/Pt multilayer films}

\author{A.\ Yagil}\affiliation{Department of Physics, Technion, Haifa 32000, Israel}
\author{A.\ Almoalem}\affiliation{Department of Physics, Technion, Haifa 32000, Israel}
\author{Anjan Soumyanarayanan}
\affiliation{Data Storage Institute, Agency for Science, Technology and Research (A*STAR), 2 Fusionopolis Way, 138634 Singapore}
\affiliation{Division of Physics and Applied Physics, School of Physical and Mathematical Sciences, Nanyang Technological University, 637371 Singapore}
\author{Anthony K.\ C.\ Tan}
\affiliation{Data Storage Institute, Agency for Science, Technology and Research (A*STAR), 2 Fusionopolis Way, 138634 Singapore}
\author{M.\ Raju}
\affiliation{Division of Physics and Applied Physics, School of Physical and Mathematical Sciences, Nanyang Technological University, 637371 Singapore}
\author{C. Panagopoulos}\email{christos@ntu.edu.sg}
\affiliation{Division of Physics and Applied Physics, School of Physical and Mathematical Sciences, Nanyang Technological University, 637371 Singapore}
\author{O.\ M.\ Auslaender}\email{ophir@physics.technion.ac.il}\affiliation{Department of Physics, Technion, Haifa 32000, Israel}

\begin{abstract}
Skyrmions are nanoscale spin configurations with topological properties that hold great promise for spintronic devices. Here, we establish their \Neel texture, helicity, and size in Ir/Fe/Co/Pt multilayer films by constructing a multipole expansion to model their stray field signatures and applying it to magnetic force microscopy (MFM) images. Furthermore, the demonstrated sensitivity to inhomogeneity in skyrmion properties, coupled with a unique capability to estimate the pinning force governing dynamics, portends broad applicability in the burgeoning field of topological spin textures.
\end{abstract}

\maketitle

The realization of nanoscale, magnetic skyrmions in metallic multilayer films has generated a surge of research \cite{Romming2013, MoreauLuchaire2016, Woo2016,Boulle2016}. 
Understanding the structure and behavior of these localized,  two-dimensional (2D) spin-textures is fundamental \cite{Nagaosa2013, Soumyanarayanan2016a, Wiesendanger2016}, with implications for spintronics technologies. 
The unique properties of skyrmions stem from their topologically non-trivial spin-configuration. The spin at the center of a skyrmion is opposite to the out-of-plane (OP) spin direction of the background [{Fig.~\ref{fig:skyrmionfield_ODE}(a-d)], and reverses over a length-scale defining the skyrmion radius ($\rsk$), which can vary from a few nanometers to microns \cite{Romming2015,Jiang2015}. 
The in-plane (IP) spin component winds chirally with helicity $\gamma$ \cite{Nagaosa2013}, ranging from \Neel \cite{Wiesendanger2016}  ($\gamma=0,\pi$) to Bloch \cite{Yu2010} ($\gamma=\pm\pi/2$) texture [Fig.~\ref{fig:skyrmionfield_ODE}(a-d)].
		
Skyrmions are generated by the anti-symmetric \DM\ interaction (DMI) found in chiral magnets \cite{Muhlbauer2009, Lee2009, Yu2010, Kiselev2011, Milde2013} and at ferromagnet/heavy-metal interfaces \cite{Fert1990, Bode2007}.  
Efforts to realize interfacial DMI have rapidly shifted from epitaxial monolayers \cite{Boulle2016} to sputtered multilayer films that host columnar room-temperature (RT) skyrmions \cite{MoreauLuchaire2016, Boulle2016, Woo2016, Nandy2016, Soumyanarayanan2017}. 
The properties of multilayer skyrmions show considerably more variation than their epitaxial counterparts. 
First, $\rsk$ can be inhomogeneous, with up to $\times2$ variations over a $\mathrm{\mu m}$-range \cite{Woo2016}. 
Next, the spin structure can evolve in all three dimensions with columnar skyrmions potentially consisting of inertial cores \cite{Woo2016, Boulle2016, Buttner2015}. 
Finally, the granularity of magnetic interactions can result in varying skyrmion configurations \cite{Soumyanarayanan2017}, which affect stability, dynamics, and switching properties  \cite{Legrand2017}. 
Any effort to understand  and exploit such skyrmions requires spatially resolved information about their static properties (e.g. size, helicity, robustness to perturbations), and an understanding of how these influence their dynamic behavior.

Here we  use magnetic force microscopy (MFM) to investigate magnetic textures in a {[}Ir(1)/Fe(0.5)/ Co(0.5)/Pt(1){]}$_\mathrm{20}$ (in parenthesis -- thickness in nm) multilayer film sputtered on a $\mathrm{SiO_2}$ substrate \cite{SuM}.  Such multilayers host RT skyrmions \cite{Soumyanarayanan2017}, which we find persist down to $T=5$~K. 
MFM is an established technique for magnetic characterization on the nanoscale, with unique, yet-untapped advantages for investigating skyrmions. 
First, MFM allows for high-resolution imaging of magnetic textures in films and devices on substrates, enabling direct comparisons with transport and thermodynamic techniques \cite{Soumyanarayanan2017}. 
Next, while MFM has been used for direct/in-situ imaging of skyrmion dynamics \cite{Hrabec2017,Legrand2017}, using it in conjunction with a quantitative physical model enables  determination of individual skyrmion properties across the disordered magnetic landscape. 
Crucially, the magnetic MFM tip, when in close proximity to skyrmions,  provides a unique window into  the response of individual skyrmions to perturbations (c.f. vortices \cite{Auslaender2009}), which may facilitate experimentally-driven modeling of mobility and switching by charge and spin currents.

Motivated by the potential of quantitative MFM, we utilize it here to investigate the characteristics of individual skyrmions. To provide an accurate physical description of the MFM signal we develop a multipole expansion for the magnetic field from skyrmions (MEFS), and fit it to our model using only two free parameters per skyrmion. 
Our fit results enable us to determine {(i)} their \Neel texture and helicity  ($|\gamma|<\pi/2$), {(ii)} to quantify $\rsk$, {(iii)} to map the spatial variation of their properties, and {(iv)} to estimate the force required to move individual skyrmions.


In this work, skyrmions were stabilized at $5$~K by finite OP  magnetic field, $\mu_0H$, after saturation at $-0.5$~T. MFM imaging was performed by rastering a magnetic tip  above the planar ($x-y$) surface of the sample. The MFM signal arises from the variation of  the tip-sample interaction force ($F_z$) induced by oscillating the height between $h$ and $h+2a$, which we track by measuring the change in resonant frequency ($\Delta f$) of the cantilever holding the tip \cite{Albrecht1991}. Such a response can be well-described provided that the cantilever motion is harmonic and that $\Delta f\ll f_{0}$, the free resonant frequency \cite{Giessibl1997}. Adapting to MFM raster scanning \cite{SuM}, the 2D Fourier transform (FT) of $\Delta f$  is related to the FT of $\partial F_{z}/\partial h$, by $\widehat{\Delta f}_{\mathbf{q}}(h)={\cal T}(qa)\widehat{\partial {F}_z/\partial h}$, 
where $q= (q_x^2+q_y^2)^{1/2}$  and ${\cal T}(qa)=-k_0^{-1}f_0I_1(qa)\exp{(-qa)}/(qa)$. Here 
$k_0\approx1$~N/m is the spring constant of the cantilever, $I_1(x)$ is a Bessel function, $f_0\approx75$~kHz, $a\approx 30$~nm, and $h\lesssim 50$~nm for sufficiently high resolution. As expected  for $a\rightarrow0$, ${\cal T}(qa) \approx -f_0/2k_0$ \cite{Albrecht1991}.


Figure~\ref{fig:data}(a) is a typical MFM image acquired at $-0.3$~T. We identify the  small round features  as skyrmions \cite{Soumyanarayanan2017}. %
As the tip and sample were polarized together, the uniformly magnetized background interacts weakly with the tip, with small variations indicating disorder \cite{Bacani2016}.
In contrast, the skyrmions, magnetized opposite to the background, display a much stronger interaction with the tip. %
The skyrmions are randomly dispersed suggesting that disorder is more important than skyrmion-skyrmion interactions under these conditions. %
The disorder reveals its role also in the zoom in Fig.~\ref{fig:data}(c), which shows that the skyrmions are not identical. 
\begin{figure}
	\centering
	\includegraphics[width=3.4in]{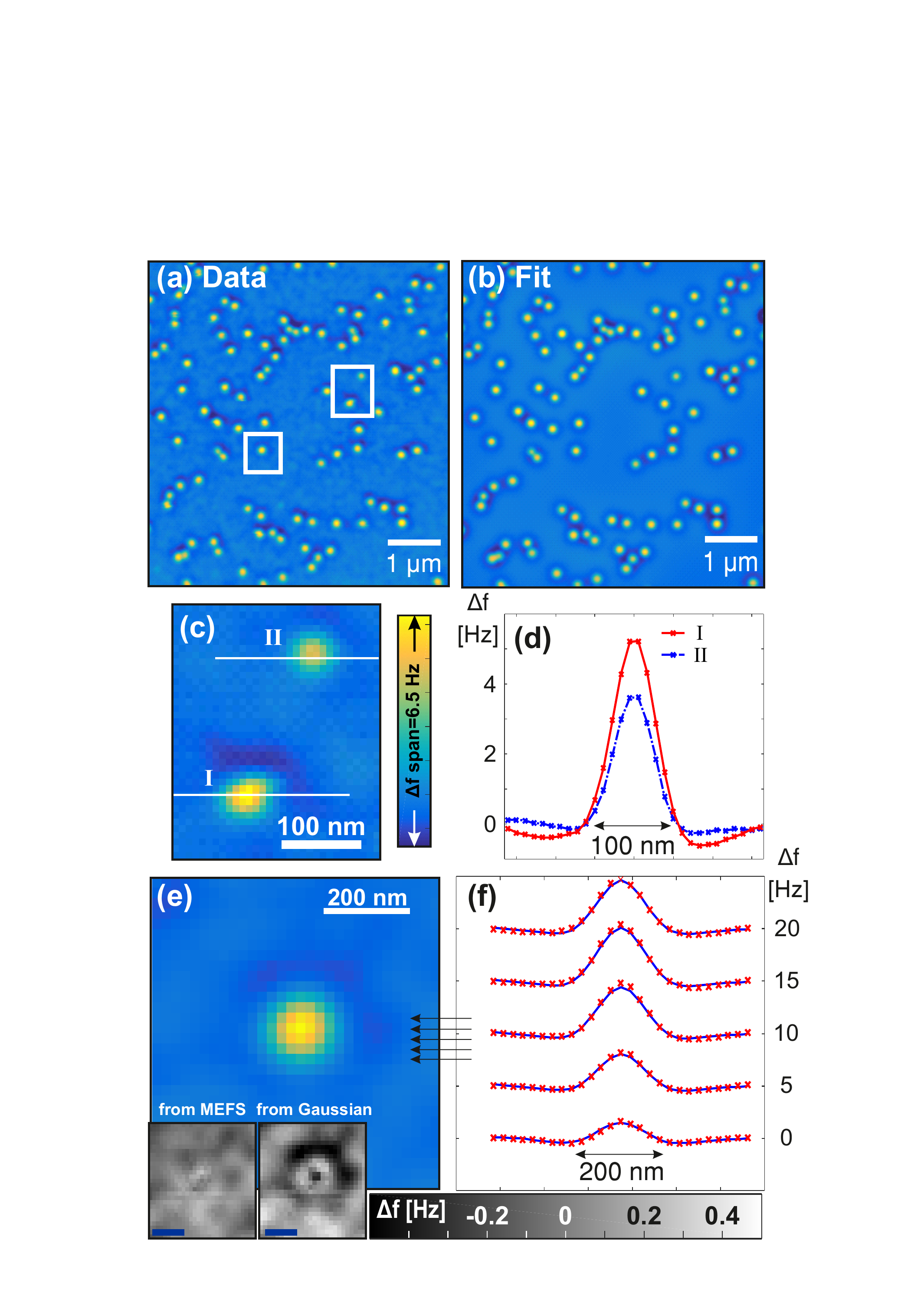}
	\caption{%
		\textbf{(a)} MFM image for $\mu_0H=-0.3$~T  with $h=50~\mathrm{nm}$. White frames show zoom areas for (c), (e).  %
		\textbf{(b)} Result of the MEFS fit assuming each skyrmion is different. %
		\textbf{(c)} Zoom on visibly different skyrmions. %
		\textbf{(d)} Line-cuts through the the two skyrmions in (c).%
		\textbf{(e)} Zoom on an individual skyrmion. Arrows indicate the positions of line-cuts in (f). \textbf{Insets:} the  difference between the data and the MEFS fit [left, detail from Fig.~\ref{fig:diff_fit}(b)], or a 2D Gaussian fit (right). [Scale bars: $200$~nm.] %
	\textbf{(f)} Line-cuts through the data in (d) offset for clarity  (x's), and the fit (lines).  }
	\label{fig:data}
\end{figure}

We now focus on understanding the signal profile of individual skyrmions [cf. Fig.~\ref{fig:data}(c),(e)].
Previously their profile has been fit to a standard line-shape, e.g. an isotropic Gaussian [cf. Fig.~\ref{fig:data}(e), right inset]. Here we present an improved framework for describing the profile [Fig.~\ref{fig:data}(e), left inset], which is physically justified from a microscopic model, more accurate, and helps unveil useful skyrmion characteristics. %

In particular, we have found that the sum of a dipolar field and a quadrupolar field describes the magnetic field of a skyrmion well [cf. fit in Fig.~\ref{fig:data}(b)]. %
Below we describe the motivation for this description, and examine the relationship between the dipole ($P_{i}$) and quadrupole ($Q_{ij}$) moments and the MFM signal.

\begin{figure}
	\centering
	\includegraphics[width=3.4in]{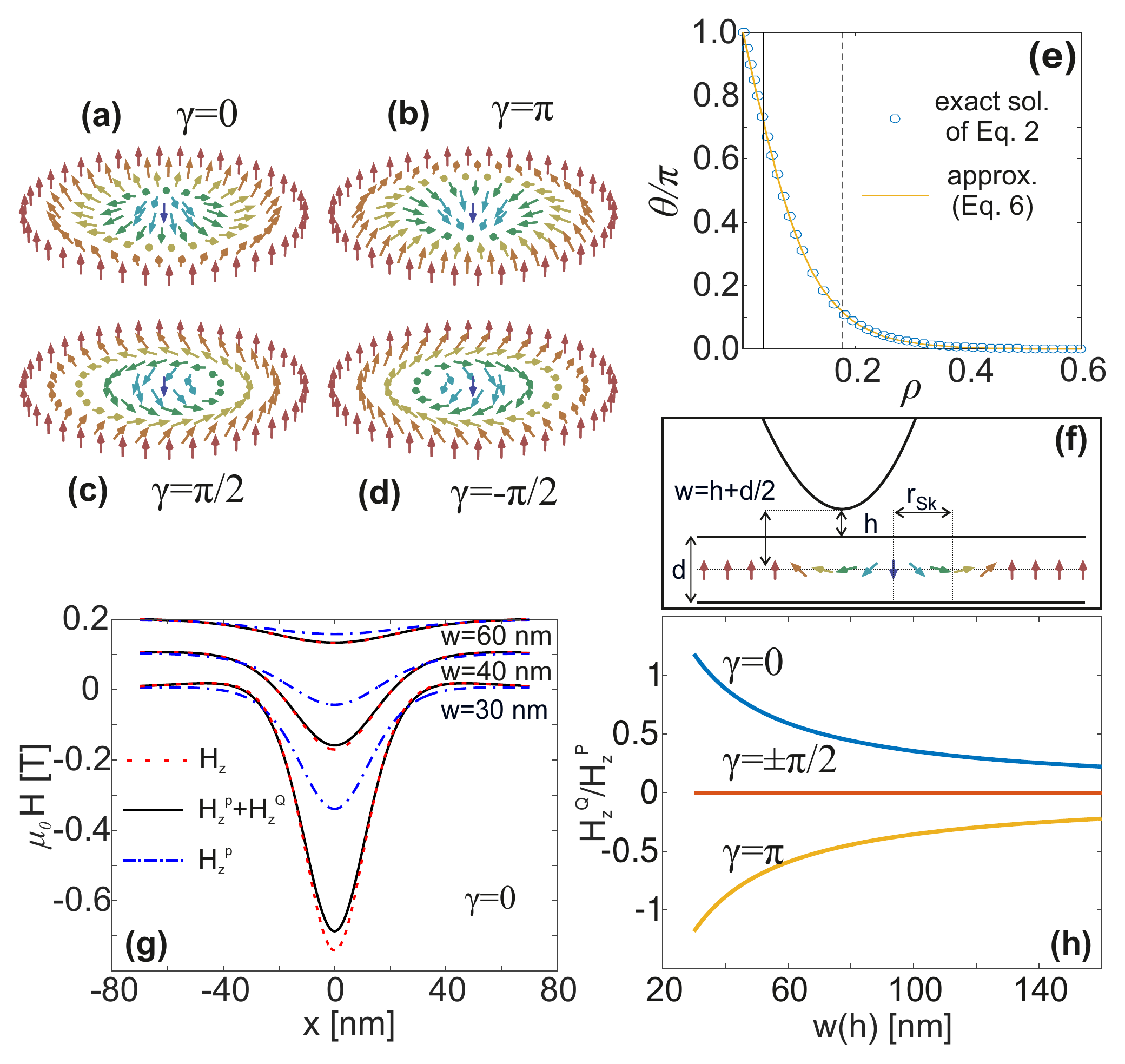}
	\caption{%
		\textbf{(a-d)} Schematic spin texture in \Neel (a,b) and Bloch (c,d) skyrmions with $m=1$ and $\gamma=0,\pi,\pm\pi/2$.
		\textbf{(e)} Plot of $\theta(\rho)$ for $k=1.536$, $b=1.474$  \cite{magpar} from a numerical solution of Eq.~\ref{eq:ODE} (circles), and a fit to Eq.~\ref{eq:romming} (line), which gives: $\sigma=0.140$,  $\rho_0=0.037$. The vertical lines show $\rho_0$ (solid) and $\rho_0+\sigma$ (dashed). $\rho=2\pi\rsk/L_D$ corresponds to $\theta/\pi=0.5$.
		\textbf{(f)} Illustration of the bottom of the tip, length-scales that we use, and a cut through a $\gamma=0$ \Neel\ skyrmion. 
		\textbf{(g)} $\mu_0H_z$, in a cut through the center of a skyrmion in a $d=20$~nm film for various values of $w=h+d/2$, with the curves offset for clarity. 
		Shown are $\mu_0H_z$ obtained from $\theta(\rho)$ in (e) as well as from the dipole and quadrupole fields   ($\equiv H^P_z$, $\equiv H^Q_z  $) from Eqs.~\ref{eq:P},~\ref{eq:Q} with $\gamma=0$.
		\textbf{(h)} $H^Q_z/H^P_z$ vs. $w$ calculated using Eqs.~\ref{eq:P},~\ref{eq:Q} for $\gamma=0,\pm \pi/2,\pi$ as a function of $w$ for $R=0$. 
	}
	\label{fig:skyrmionfield_ODE}
\end{figure}


The magnetic field generated by the skyrmion magnetization ($\mathbf{M}$) determines its MFM signature. 
For a uniformly magnetized thin film hosting an axially-symmetric skyrmion with vorticity $m$ magnetized along $\pm\hat{z}$, $\mathbf{M}$ is given by \cite{Nagaosa2013}:
$\mathbf{M}(\rho,z)/M_s(z)=\sin\theta(\rho)\cos\psi(\varphi)\hat{x}+\sin\theta(\rho)\sin\psi(\varphi)\hat{y}+
\left[\pm1+\cos\theta(\rho)\right]\hat{z}$. 
Here $\psi(\varphi)=m\varphi+\gamma$, where $\varphi$ is the axial angle, $\theta$ is the polar angle, $\rho=2\pi r/L_{D}$ ($r$ is the distance from the skyrmion center, $L_{D}\equiv4\pi A/|D|$ is the domain wall thickness, $A$ is the exchange stiffness and $D$ is the micromagnetic DMI strength). Meanwhile $M_{s}(z)=M_{s}^{0}\left[\Theta({z+d/2})-\Theta({z-d/2})\right]$, where $M_{s}^{0}$ is the saturation magnetization, $d$ is the film thickness, and $\Theta(z)$ is the Heaviside function. $\theta(\rho)$ is a solution to well-known ordinary differential equations \cite{Leonov2016a} with appropriate boundary conditions. For $\theta(0)=\pi$, $\theta(\infty)=0$, and $m=1$ we have \cite{Nagaosa2013, Soumyanarayanan2016a}:
\begin{equation}\label{eq:ODE}
\theta'' + \frac{\theta'+2\sin^2\!\theta}{\rho}-\sin\!\theta\cos\!\theta\left({\rho}^{-2}+k\right) -b\sin\!\theta=0.
\end{equation}
Here $b=B/B_D$ ($B_D\equiv D^2/2AM_s^0$), and  $k=4AK/D^2$, where $K$ is the effective anisotropy. Fig.~\ref{fig:skyrmionfield_ODE}(e) shows a solution for parameters typical to multilayers \cite{magpar}.



The magnetic field from a localized magnetic structure can be described by a multipole expansion \cite{Jackson1998}.  %
For this purpose we define a magnetic scalar potential $\mathbf{H}=-\nabla\Phi$ \cite{SuM}.  %
The first term of the resulting MEFS is proportional to  $P_{i}\equiv-\int r_{i}{\mathbf{\nabla}\cdot\mathbf{M}}\,dv$, the second to  $Q_{ij}\equiv-\int\left(3r_{i}r_{j}-\mathbf{r}^{2}\delta_{ij}\right){\mathbf{\nabla}\cdot\mathbf{M}}\,dv$. For axially symmetric skyrmions with $m=1$, $\mathbf{P}=P\hat{z}$ and $Q_{ij}$ is diagonal with $Q_{xx}=Q_{yy}=-Q_{zz}/2\equiv Q$. Thus:%
\begin{equation}\label{eq:multipole}
\Phi\left(r,h\right)\approx\frac{1}{4\pi}\left(\frac{Pw(h)}{\left[r^2+w(h)^2\right]^{3/2}}+\frac{Q}{2}\frac{r^2-2w(h)^2}{\left[r^2+w(h)^2\right]^{5/2}}\right),
\end{equation}
where $w(h)\equiv h+d/2$  and \cite{SuM}:
\begin{eqnarray}
\label{eq:P} P &=&
2\pi M_s^0d\left(\frac{L_D}{2\pi}\right)^2\int_0^\infty\!\!\! d\rho\rho \left\{\pm1+\cos\left[\theta(\rho)\right]\right\}, \\
\label{eq:Q} Q&=&2\pi M_s^0d\left(\frac{L_D}{2\pi}\right)^3\cos{\gamma}\int_0^\infty\!\!\!d\rho\rho^2\sin\left[\theta(\rho)\right].
\end{eqnarray}
The sign of $P$ corresponds to the OP magnetization of the skyrmion ($\pm\hat{z}$). The sign of $Q$ indicates whether the IP magnetization points away ($+$) or towards ($-$) the center, thus determining the helicity of \Neel skyrmions, which is difficult to extract from other techniques \cite{Pulecio2016}. Importantly, for Bloch skyrmions $Q=0$. 

To estimate $P$ and $Q$, we approximate the solution of Eq.~\ref{eq:ODE} \cite{Romming2015}:
\begin{equation}\label{eq:romming}
\pi-\theta(\rho)\approx\sin^{-1}\left[\tanh\left(\eta_+\right)\right]+\sin^{-1}\left[\tanh\left(\eta_-\right)\right],
\end{equation}
where $\eta_\pm\equiv(\rho\pm\rho_0)/\sigma$, $\sigma$ parameterizes the  domain wall thickness and $\rho_0$ the skyrmion radius. For $\rho_0/\sigma\ll1$  $Q/|P|\approx\rsk\cos\gamma$, where $M_z(r=\rsk)=0$ \cite{SuM}.
For Fig.~\ref{fig:skyrmionfield_ODE}(e)  a fit to Eq.~\ref{eq:romming} gives $\sigma=0.140$, $\rho_0=0.037$.

The analytical model is substantiated by numerical calculations of the magnetic field from a $\gamma=0$ skyrmion in a film with $d=20$~nm, as illustrated in Fig.~\ref{fig:skyrmionfield_ODE}(f). %
A comparison between the exact solution and MEFS, shown in Fig.~\ref{fig:skyrmionfield_ODE}(g), suggests that Eq.~\ref{eq:multipole} describes the stray field very well with the quadrupolar contribution increasing gradually as $w(h)$ is reduced. %
Figure~\ref{fig:skyrmionfield_ODE}(h) shows the ratio between the quadrupole ($H_z^Q$) and the dipole ($H_z^P$) contributions to $H_z$,  from Eqs.~\ref{eq:multipole}-\ref{eq:Q} with $\gamma=0,\pm\pi/2,\pi$. %
As expected, for Bloch skyrmions ($\gamma=\pm\pi/2$) $Q=0$, and the sign of $Q$ is opposite for the two kinds of \Neel skyrmions ($\gamma=0,\pi$). Therefore $Q/P$ allows the direct determination of skyrmion helicity.


To fit the MFM data using Eq.~\ref{eq:multipole}, we model the tip as a thin shell with axial symmetry (\cite{SuM}), and illustrated in Fig.~\ref{fig:skyrmionfield_ODE})}: $M^t_z\!(\mathbf{R},z)=m_0 \delta\left[|\mathbf{R}|-g(z)\right].$
Here $m_{0}=M_{0}\,t$, where $M_{0}$ is the tip magnetization and $t$ is the thickness of the magnetic coating; $z$ is along the tip axis and $g(z)$ is the radius of the tip in a constant-$z$ cut.  Given $g(z)$, Eq.~\ref{eq:multipole} implies \cite{SuM}:
\begin{eqnarray}\label{eq: wide tip int}
\frac{\partial F_z}{\partial h}=&&-C\!\!\int_0^\infty\!\!\!\!\!\!\! dq \!\! \int_0^\infty\!\!\!\!\!\!\! dz q^4\left(1-q\frac{Q}{2P}\right)
J_0[g(z)q]g(z)J_0(rq)\nonumber\\ &&\;\;\;\;\;\;\;\;\;\;\;\;\;\;\;\;\;\;\;\;\;\;\;\;\;\;\;\;\;\;\;\;\;\;\;\;\;\;\;\;\;\;\times e^{-q\left[w(h)+z\right]},
\end{eqnarray}
where we have used the FT of $\Phi\left(r,h=0\right)$,  $J_0(x)$ is a Bessel function, and $C$ is a constant proportional to $P$ that determines the scale of the skyrmion-tip interaction. %
A fit  using Eq.~\ref{eq: wide tip int} is computationally expensive. Therefore, we first determine the skyrmion positions, peak magnitude ($\Delta f_\mathrm{max}$)  and full-width-at-half-maximum (FWHM) by fitting to a simplified model of the tip \cite{SuM}.

\begin{figure}
	\centering
	\includegraphics[width=3.4in]{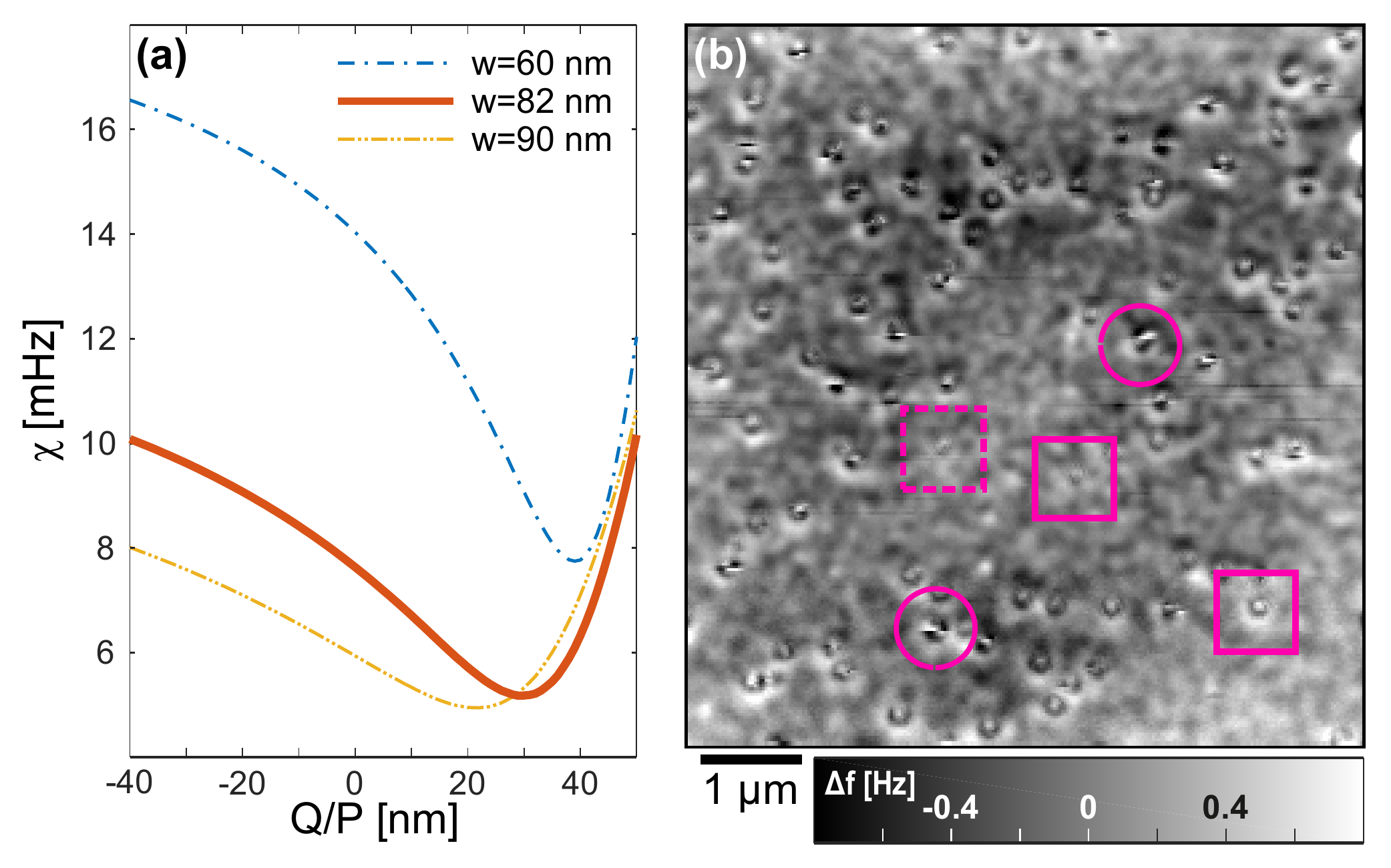}
	\caption{%
		\textbf{(a)} Plot of $\chi(Q/P)$ for a representative skyrmion for the measured $w(h)=82$~nm. The other curves show the influence of $w$. For each $Q/P$ we chose $C$ that minimizes $\chi$ in the area marked by the dashed rectangle in (b) after subtracting the contributions of all other skyrmions, which were fit using a simpler model \cite{SuM}. 
		\textbf{(b)}  The difference between data and fit in Fig.~\ref{fig:data}(a),(b). The dashed rectangle shows a $650$~nm square, where $\chi$ is calculated for a representative skyrmion [in (a)]. 
		The location of representative skyrmions showing discontinuous (circles) or negligible (rectangles) residual are highlighted. 
	}
	\label{fig:diff_fit}
\end{figure}

Next, we fit the signal from the skyrmions using  a more accurate model for the tip: $g(z)=\alpha (z^4+\beta z)^{{1}/{4}}$, with $\alpha=0.24$ and $\beta=2.7\cdot10^{6}$~$\mathrm{nm}^{3}$ \cite{SuM}. %
This fit includes only two free parameters per skyrmion ($C$ and $Q/P$), as the positions are set from the fit to the simplified tip model, $\alpha$ and $\beta$ are determined from scanning electron microscopy, and we measure $d$ and $h$.
Figure~\ref{fig:diff_fit}(a) shows the dependence of $\chi$ (root-mean-square of the error) on $Q/P$ for a representative skyrmion for three values of $w$, including the actual value for the data in Fig.\ref{fig:data}(a), $w(h)=82$~nm. This value includes $2$~nm for a capping-layer, and the $d=60$~nm magnetic part of the stack \cite{SuM}. For all skyrmions we find a single global minimum corresponding to the optimal $Q/P$, that becomes shallower for larger $w(h)$. As expected, $h$ and $d$ have a direct impact on how precisely $Q/P$ can be determined. Based on such analysis we conclude that $Q/P\gtrsim0$ for the skyrmions in our film, indicating \Neel texture.%

Figure~\ref{fig:data}(b) shows the fit and  Fig.~\ref{fig:diff_fit}(b) the residual we obtain upon repeating this fitting procedure for all skyrmions in Fig.~\ref{fig:diff_fit}(a). This reveals several subtle features. %
First, the nanoscale variations in the background that are typical of the inhomogeneous magnetic structure of sputtered multilayer films \cite{Bacani2016}. %
Second, are discontinuities for some skyrmions [e.g. circles in Fig.~\ref{fig:diff_fit}(b)], likely due to MFM tip-induced skyrmion motion. Other explanations, such as irregular skyrmion shapes, cannot give such sharp fit residuals.  %
These observations, in conjunction with the variability in skyrmion properties, are  direct consequences of inhomogeneous magnetic interactions \cite{Bacani2016}, and reinforce the need for individual fit parameters to accurately describe multilayer skyrmions. 

\begin{figure}
	\centering
	\includegraphics[width=3.4in]{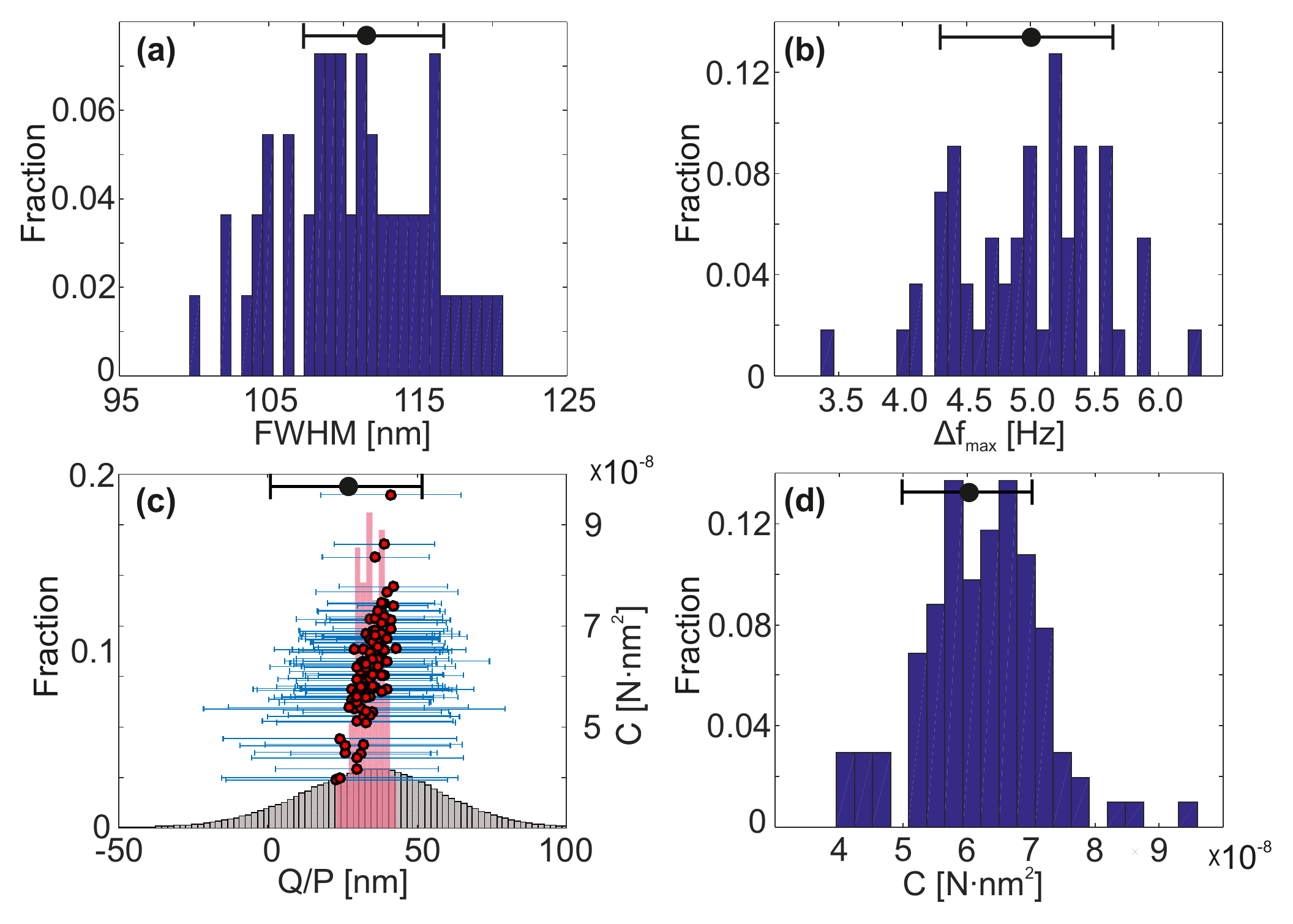}
	\caption{Histograms and plots of skyrmion parameters from the fit to Fig.~\ref{fig:data}(a). Dots: mean, horizontal bars: $70\%$ confidence intervals. %
		\textbf{(a,b)} FWHM and $\Delta f_\mathrm{max}$.
		\textbf{(c)} Histograms of $Q/P$ and a plot of $C$ vs. $Q/P$. The large error bars reflect the proximity of the resolution limit. The narrow distribution in (c) is directly from the fit, while the wide distribution is derived by accounting for the large error bars by assuming they are normally distributed. %
		\textbf{(d)} $C$. %
		[The data in (a)-(d) was taken from $91$ out of $104$ skyrmions which did not exhibit residual discontinuities, cf. Fig.~\ref{fig:diff_fit}(b)].  
	}
	\label{fig:indi_fit_hist}
\end{figure}

Figure~\ref{fig:indi_fit_hist} shows histograms of the individual skyrmion parameters we obtain from the fit to Fig.~\ref{fig:data}(a). %
Figure~\ref{fig:indi_fit_hist}(a) shows that FWHM, which includes tip effects, varies by $\sim20\%$ ($100-120$~nm). Its larger magnitude compared to RT values for similar films \cite{Soumyanarayanan2017} is likely due to the changed magnetic parameters at $5$~K. %
Notably however, the uniform FWHM [Fig.~\ref{fig:indi_fit_hist}(a)] is in contrast to the considerable variability of $\Delta f_\mathrm{max}$ [Fig.~\ref{fig:indi_fit_hist}(b)], indicating a significant variation in the stray field strength of the skyrmions. %

The model allows us to go beyond conventional MFM to extract a typical length scale ($Q/P$) that, unlike FWHM, is dissociated from both the tip shape and the effect of $h$, and can therefore be lower. This ability to extract information on true length scales indicates the power of MEFS. %
Figure~\ref{fig:indi_fit_hist}(c) shows histograms for $Q/P$. The narrow histogram is for the actual fit results, but as the fit uncertainty is large [cf. $C$ vs. $Q/P$ and Fig.~\ref{fig:diff_fit}(a)], we generated the wider histogram. For this we  assumed that each value of $Q/P$ is drawn from a normal distribution $\mathcal{N}(Q/P,\delta\frac{Q}{P})$ with the width $\delta\frac{Q}{P}$ given by the error shown in the $C(Q/P)$ plot. %
We conclude that {$Q/P\approx34$~nm} (standard deviation $27$~nm). %
This number represents the shape of skyrmions that do not exhibit discontinuities. By comparing to scans with larger $h$ we conclude that changes induced by the field from the tip are too subtle for us to observe. Integrating the wide histogram we find that $Q/P>0$ with probability $0.9$. %
This likely rules out Bloch skyrmions and implies that our skyrmions have \Neel texture with helicity $|\gamma|<\pi/2$. While this is consistent with \Neel skyrmions with helicity $\gamma=0$ [Fig.~\ref{fig:skyrmionfield_ODE}(a),(f)], we cannot rule out the presence of a partial Bloch component \cite{Rowland2016}.

Figures~\ref{fig:indi_fit_hist}(c),(d) show no correlation between $C$ and $Q/P$, and that the relative spread of $C$ is smaller ($C=61\pm9~\mathrm{nN\cdot nm^2}$). %
The contrasting spreads are likely due to the inherent sensitivity of $C$ to $P$, rather than to $Q/P$. $Q/P$ provides finer information on the skyrmion structure \cite{SuM}, and is therefore more sensitive to disorder, which in turn contributes to its spread. Crucially, a key utility of our model is the ability to calculate the force exerted by the tip on skyrmions. We estimate that a skyrmion with the mean $Q/P$ and $C$ experiences a lateral force of $F_{tip}\approx1$~pN as a result of interaction with the MFM tip \cite{SuM}. As this force was sufficient to move only some of the skyrmions, we estimate $F_\mathrm{pin}\approx F_\mathrm{tip}$, where $F_\mathrm{pin}$ is the typical force required to move a skyrmion. Using the Lorentz force to estimate a critical current for adiabatic manipulation of skyrmions, we obtain \cite{Lin2013} $J=(F_\mathrm{pin}/d)/(h/e)\approx4\cdot10^{9}\mathrm{A/m^2}$. This $5$~K value is smaller than RT values reported previously on similar samples \cite{Woo2016}, and indicates that accounting for non-adiabatic processes and interlayer interactions may provide an improved estimate for bottom-up predictive modeling of skyrmion dynamics \cite{Finocchio2016}.


In summary, we have shown that MFM images of skyrmions can be quantitatively reproduced by modeling the magnetic field from a skyrmion using a closed expression from a multipole expansion with two free parameters per skyrmion, with several conclusions.  %
First, based on $|Q/P|\gtrsim0$ we can rule out with $\approx90\%$ certainty the skyrmions in our Ir/Fe/Co/Pt multilayers as purely Bloch textured.  %
The sign of $Q/P$ independently establishes that these skyrmions are \Neel textured with helicity $|\gamma|<\pi/2$, consistent with micromagnetic calculations \cite{Soumyanarayanan2017}. %
Second, the magnitude of $Q/P$ provides the estimate $\rsk=36\pm28$~nm for $\gamma=0$. %
Third, the spread of $\rsk$ and the $\Delta f_\mathrm{max}$ can be directly used to estimate the corresponding inhomogeneity of magnetic interactions. %
In particular, $\Delta f_\mathrm{max}$ [Fig.~\ref{fig:indi_fit_hist}(b)] is expected to be sensitive to variations in $D$ \cite{Bacani2016}, and $\rsk$ is expected to be sensitive to variations in $\Keff$ \cite{Kim2017}.  %
Fourth, we have estimated the pinning force skyrmions experience, and the critical current density for skyrmion motion.
Finally, the utility of the physical analysis we presented beyond MFM, the compatibility with device configurations, and the relative computational simplicity that allows to apply it easily to large arrays of skyrmions, all bode well for future use towards both applications and basic science.


\begin{acknowledgments}
	We are grateful for input from D. Arovas, K. Kuchuk, Shi-Zeng Lin, D. Podolsky, Y. Shechner, I. Schlesinger, U. Sivan, and A. Turner. The work in Technion was supported by the Israel Science Foundation (Grant no. 1897/14). The work in Singapore	was supported by the Ministry of Education (MoE) -- Academic Research Fund (Ref. No. MOE2014-T2-1-050), the National Research Foundation	-- NRF Investigatorship (Reference No. NRF-NRFI2015-04), and the A{*}STAR Pharos Fund (1527400026). We would also like to	thank the Micro Nano Fabrication Unit at the Technion.
\end{acknowledgments}


\begin{thebibliography}{34}
\bibitem [{mag()}]{magpar}%
  \BibitemOpen
  \href@noop {} {}\bibinfo {note} {Reasonable parameters for multilayer films
  are $A=12$~pJ/m, $D=2.5$~mJ/m$^{2}$, $K=0.20$~MJ/m$^{3}$, $\mu_{0}H=0.38$~T
  and $M_{s}=1.02$~MA/m. In this case $L_{D}=60$~nm, $K_{0}=0.13$~~MJ/m$^{3}$,
  $B_{D}=0.26$~T and consequently $k=1.54$ and $b=1.47$.}\BibitemShut {Stop}%
\end{thebibliography}

\end{document}